\documentclass[prl,twocolumn,superscriptaddress,showpacs]{revtex4}

\usepackage{graphicx}
\usepackage{amssymb}
\usepackage{amsmath}
\usepackage{amsfonts}
\usepackage{mathrsfs}
\usepackage[usenames]{color}
\definecolor{mygrey}{gray}{0.75}

\newcommand {\be} {\begin{equation}}   
\newcommand {\ee} {\end{equation}}
\newcommand {\bea} {\begin{eqnarray}}
\newcommand {\eea} {\end{eqnarray}}
\newcommand {\bes} {\begin{displaymath}}
\newcommand {\ees} {\end{displaymath}}
\newcommand {\beas} {\begin{eqnarray*}}
\newcommand {\eeas} {\end{eqnarray*}}

\begin{document}

\title{Sequence sensitivity of breathing dynamics in heteropolymer DNA}

\author{Tobias Ambj{\"o}rnsson}
\affiliation{NORDITA, Blegdamsvej 17, DK-2100 Copenhagen {\O}}
\author{Suman K. Banik}
\affiliation{Dept. of Physics, Virginia Polytechnic Institute
 and State University, Blacksburg, VA 24061-0435, USA}
\author{Oleg Krichevsky}
\affiliation{Physics Department, Ben Gurion University, Be'er Sheva 84105,
Israel}
\author{Ralf Metzler}
\affiliation{NORDITA, Blegdamsvej 17, DK-2100 Copenhagen {\O}}

\begin{abstract}
We study the fluctuation dynamics of localized denaturation bubbles in
heteropolymer DNA with a master equation and complementary stochastic
simulation based on novel DNA stability data. A significant dependence
of opening probability and waiting time between bubble events on the local
DNA sequence is revealed and quantified for a biological sequence of the T7
bacteriophage. Quantitative agreement with data from fluorescence
correlation spectroscopy (FCS) is demonstrated.
\end{abstract}


\maketitle

The biological function of DNA largely relies on its physical properties:
Protein binding is sensitive to local DNA structure \cite{alberts}, DNA
looping facilitates the search of binding proteins for their specific site
\cite{michael}, and DNA knots impair transcription or act as barriers
between different genome regions \cite{knots}. Similarly, local
denaturation of DNA is necessary for protein binding to DNA single-strand
\cite{mark,mark1,tobias}, and is implicated in transcription initiation
\cite{kalo,yeramian}. DNA melting has a long tradition in statistical physics
\cite{poland}. Its biological relevance is due to the fact that the free 
energy for breaking a single base pair (bp) at physiological temperature
is $\sim k_BT$ \cite{blake,FK}. Renewed interest in DNA melting, from a
physics perspective is nourished by the possibility to measure the fluctuation
\emph{dynamics\/} of local denaturation bubbles by single molecule FCS
\cite{oleg}.

We present a master equation (ME) and complementary stochastic
simulation, that provides the time series of the bubble fluctuations.
A full two-variable formulation in terms of bubble size $m$ and left fork
location $x_L$ allows to investigate an arbitrary sequence of bps, beyond
previous homopolymer \cite{tobias} and random energy models \cite{hwa}. In
certain limits, the ME can be solved analytically.
We employ DNA stability data from a novel approach
measuring the ten stacking interactions separately and, inter alia predicting
a distinct asymmetry between AT/AT and AT/TA nearest neighbour bps \cite{FK}.
As proved on recent FCS experimental data our model describes well the
bubble dynamics with only one free parameter. We demonstrate the delicate
sensitivity of bubble dynamics to the local sequence of heterogeneous DNA on
the promoter sequence of the T7 bacteriophage, and illustrate good potential
for nanosensor applications.

\begin{figure}
\begin{center}
\includegraphics[width=6.8cm]{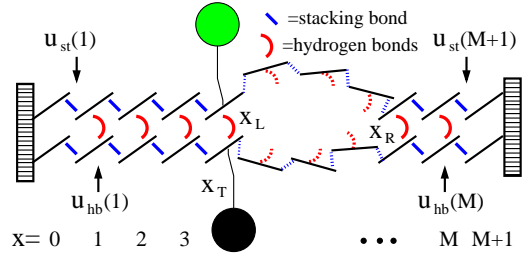}
\end{center}
\caption{Clamped bubble domain with internal bps $x=1$ to $M$, statistical
weights $u_{\rm hb}(x)$, $u_{\rm st}(x)$, and tag position $x_T$.}
\label{fig:bubbles}
\end{figure}

\emph{Model.}~With typical experimental setups \cite{oleg} in mind, we
consider a segment of double-stranded DNA with $M$ internal bps, that are
clamped at both ends (Fig.~\ref{fig:bubbles}). The full sequence of bps
enters via the position-dependence of the statistical weights $u_{\rm
hb}(x)=\exp\{\epsilon_{\rm hb}(x)/[k_BT]\}$ for breaking the hydrogen-bonds
of the bp at position $x$, and $u_{\rm st}(x)=\exp\{\epsilon_{\rm st}(x)/
[k_BT]\}$ for disrupting the stacking interactions between bps $x-1$ and
$x$. Due to the high free energy barrier for bubble initiation ($\xi\ll 1$),
opening and merging of multiple bubbles are rare events, such that a
one-bubble description is appropriate. The positions $x_L$ and $x_R$ of the
zipper forks correspond to the right- and leftmost closed bp of the
bubble. $x_L$ and $x_R$ are stochastic variables, whose time evolution in
the energy landscape defined by the partition factor ($m\ge 1$)
%these are stochastic variables whose time evolution characterizes
%the bubble dynamics. In terms of $x_L$ and bubble size $m=x_R-x_L-1$,
%the bubble partition factor is ($m\ge 1$)
\be
\mathscr{Z}(x_L,m)=\frac{\xi'}{(1+m)^c}
\prod_{x=x_L+1}^{x_L+m} \hspace*{-0.2cm}u_{\rm hb}(x)
\prod_{x=x_L+1}^{x_L+m+1}\hspace*{-0.2cm}u_{\rm st}(x)
\label{part}
\ee
characterizes the bubble dynamics. $\mathscr{Z}$ is written in terms of $x_L$
and bubble size $m=x_R-x_L-1$, with $\mathscr{Z}(m=0)=1$. Here, $\xi'=2^c\xi$,
where $\xi\approx 10^{-3}$ is the ring factor for
bubble initiation from Ref.~\cite{FK} that is related to the cooperativity
parameter $\sigma_0\approx 10^{-5}$ \cite{poland,blake} by $\sigma_0=\xi
\exp\{\epsilon_{\mathrm{st}}\}$ \cite{FK}. For the entropy loss on forming a
closed polymer loop we assign the factor $(1+m)^{-c}$ \cite{blake,fixman} and
take $c=1.76$ for the critical exponent \cite{richard}. Note that a bubble
with $m$ open bps requires breaking of $m$ hydrogen bonds and $m+1$ stacking
interactions.

The zipper forks move stepwise $x_{L/R}\rightarrow x_{L/R}\pm 1$ with rates
$\mathsf{t}^{\pm}_{L/R}(x_L,m)$. We define for bubble size decrease
\be
\label{eq:t_L_plus}
\mathsf{t}^+_L(x_L,m)=\mathsf{t}^-_R(x_L,m)=k/2 \qquad (m\ge 2)
\ee
for the two forks \cite{REM}. The rate $k$ characterizes a single bp zipping.
Its independence of $x$ corresponds to the view that bp closure requires the
diffusional encounter of the two bases and bond formation; as sterically AT
and GC bps are very similar, $k$ should not significantly vary with bp stacking.
$k$ is the only adjustable parameter of our model, and has to be determined
from experiment or future MD simulations. The factor $1/2$ is introduced for
consistency \cite{tobias}. Bubble size increase is controlled by
\begin{eqnarray}
\nonumber
\mathsf{t}_{L}^{-}(x_L,m)&=&ku_{\rm st}(x_L) u_{\rm hb}(x_{L})
s(m)/2,
\label{eq:t_L_minus}\\
\mathsf{t}_{R}^{+}(x_L,m)&=&ku_{\rm st}(x_R+1) u_{\rm hb}(x_{R})
s(m)/2,
\label{eq:t_R_plus}
\end{eqnarray}
for $m\ge 1$, where $s(m)=\{(1+m)/(2+m)\}^c$. Finally, bubble initiation and
annihilation from and to the zero-bubble ground state, $m=0 \leftrightarrow
1$ occur with rates
\bea
\nonumber
\mathsf{t}_G^+(x_L)&=&k\xi's(0)u_{\rm st}(x_L+1)u_{\rm hb}(x_L+1)u_{\rm
st}(x_L+2)\\
\mathsf{t}_G^-(x_L)&=&k.
\label{eq:t_G_plus}
\eea
The rates $\mathsf{t}$ fulfill detailed balance conditions. The annihilation
rate $\mathsf{t}_G^-(x_L)$ is twice the zipping rate of a single fork, since
the last open bp can close either from the left or right. Due to the clamping,
$x_L\ge 0$ and $x_R\le M+1$, ensured by reflecting conditions $\mathsf{t}
_L^-(0,m)=\mathsf{t}_R^+(x_L,M-x_L)=0$. The rates $\mathsf{t}$ together
with the boundary conditions fully determine the bubble dynamics.

\begin{figure}
\begin{center}
\includegraphics[width=8.8cm]{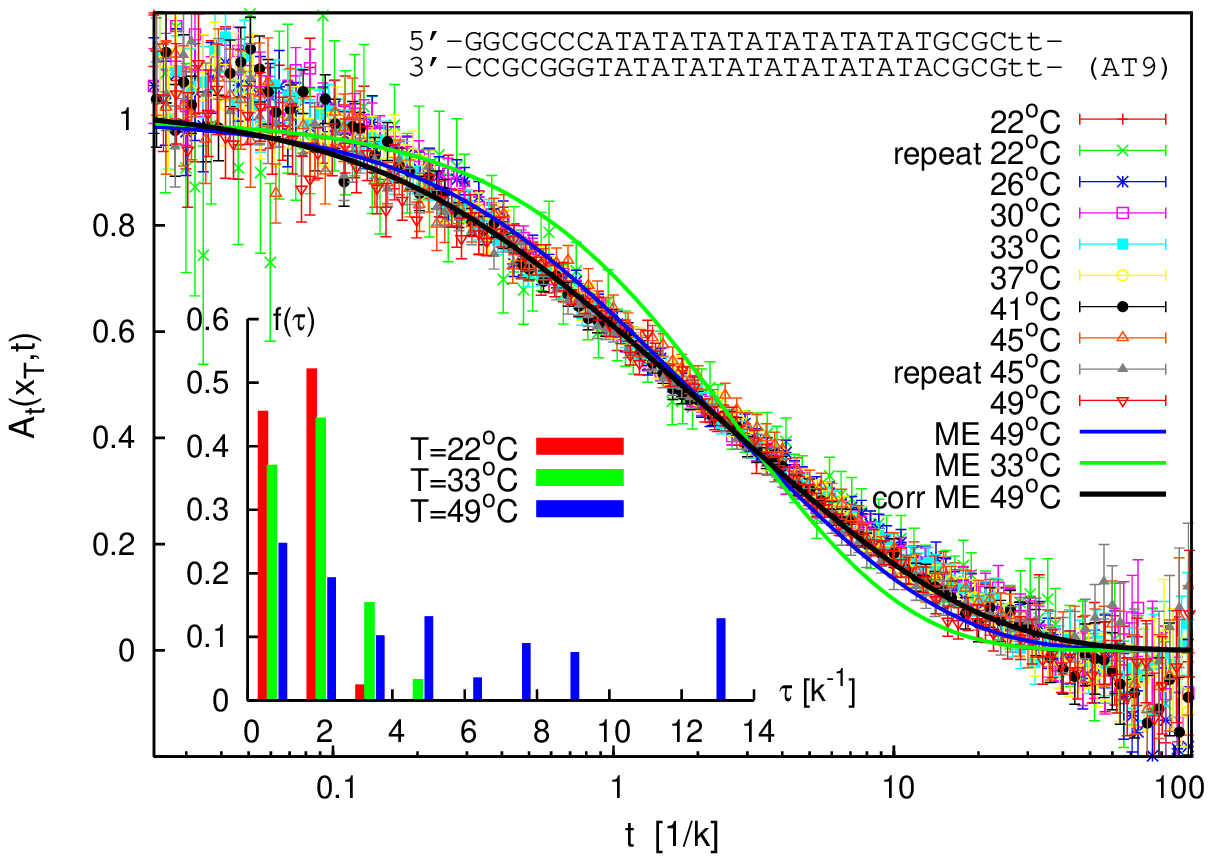}
\end{center}
\caption{Scaling plot of $A_t(x_T,t)$ at various $T$ for the sequence
AT9 from \cite{oleg}. Inset: Relaxation time spectrum.
See text.}
\label{autocorr}
\end{figure}

In the FCS experiment fluorescence occurs if the bps in a
$\Delta$-neighbourhood of the fluorophore position $x_T$ are open
\cite{oleg}. Measured fluorescence time series thus correspond to
the stochastic variable $I(t)$, that takes the value 1 if at least all bps
in $[x_T-\Delta,x_T+\Delta]$ are open, else it is 0. The time averaged
($\overline{\,\,\cdot\,\,}$) fluorescence autocorrelation
\begin{equation}
\label{autocf}
A_t(x_T,t)=\overline{I(t)I(0)}-\overline{I(t)}^2
\end{equation}
for the sequence AT9 from \cite{oleg} are rescaled in Fig.~\ref{autocorr}
\cite{remm}.

\emph{ME.}~DNA breathing is described by the probability
distribution $P(x_L,m,t)$ to find a bubble of size $m$ located at $x_L$
whose time evolution follows the ME $\partial P(x_L,m,t)/\partial
t=\mathbb{W} P(x_L,m,t)$. The transfer matrix $\mathbb{W}$ incorporates
the rates $\mathsf{t}$. Detailed balance guarantees equilibration toward
$\lim_{t\to\infty}P(x_L,m,t)=\mathscr{Z}(x_L,m)/\mathscr{Z}$, with
$\mathscr{Z}=\sum_{x_L,m}\mathscr{Z}(x_L,m)$ \cite{tobias,vankampen}. The
ME and the explicit construction of $\mathbb{W}$ are discussed at
length in Refs.~\cite{tobias,unp}. Eigenmode analysis and matrix
diagonalization produces all quantities of interest such as the ensemble
averaged autocorrelation function
\begin{equation}
\label{auto}
A(x_T,t)=\langle I(t)I(0)\rangle-(\langle I\rangle)^2.
\end{equation}
$\langle I(t)I(0)\rangle$ is proportional to the survival density that the
bp is open at $t$ and that it was open initially \cite{tobias,unp}.

In Fig.~\ref{autocorr} the blue curve shows the predicted behaviour of $A(
x_T,t)$, calculated for $T=49^\circ$C with the parameters from
\cite{FK}. As in the experiment we assumed that fluorophore and quencher
attach to bps $x_T$ and $x_T+1$, that both are required open to produce
a fluorescence signal. From the scaling plot, we calibrate the zipping
rate as $k=7.1\times 10^4/$s, in good agreement with the findings from
Ref.~\cite{oleg,REM1}. The calculated behaviour reproduces the data within
the error bars, while the model prediction at $T=35^\circ$C shows more
pronounced deviation. Potential causes are destabilizing effects of the
fluorophore and quencher, and additional modes that broaden the decay
of the autocorrelation. The latter is underlined by the fact that for
lower temperatures the relaxation time distribution $f(\tau)$, defined
by $A(x_T,t)=\int\exp(-t/\tau)f(\tau)d\tau$, becomes narrower
(Fig.~\ref{autocorr} inset).
Deviations may also be associated with the correction for diffusional motion
of the DNA construct, measured without quencher and neglecting
contributions from internal dynamics \cite{oleg1}. Indeed, the black
curve shown in Fig.~\ref{autocorr} was obtained by a 3\% reduction of the
diffusion time \cite{REM2}; see details in \cite{unp}.

\begin{figure}
\begin{center}
\includegraphics[width=8.8cm]{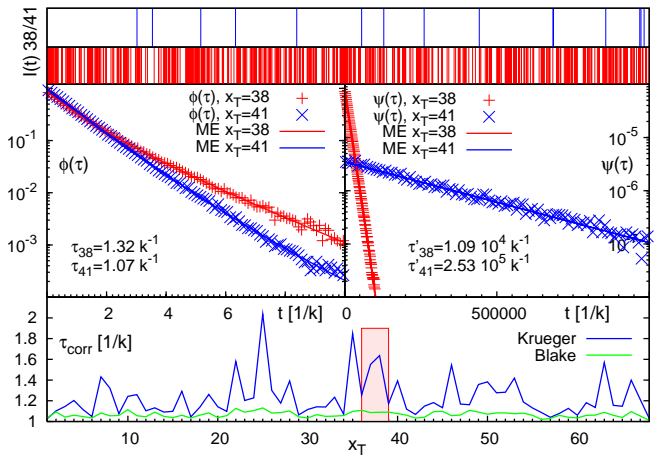}
\end{center}
\caption{Time series $I(t)$ for the T7 promoter, with
$x_T=38$, 41. Middle: Waiting time ($\psi(\tau)$) and fluorescence
time ($\phi(\tau)$) densities. Bottom: Mean fluorescence
time for $\Delta=0$.}
\label{signal}
\end{figure}

\emph{Stochastic simulation.}~Based on the rates $\mathsf{t}$, stochastic
simulations give access to single bubble fluctuations
\cite{suman,unp}. Our customized Gillespie algorithm uses
the joint probability density of waiting time $\tau$ and path $\mu=+/-$,
\begin{equation}
\label{gill}
P(\tau,\mu,\nu)=\mathsf{t}^{\mu}_{\nu}(x_L,m)\exp\left(-\tau\sum_{\mu,\nu}
\mathsf{t}^{\mu}_{\nu}(x_L,m)\right),
\end{equation}
defining for given state $(x_L,m)$ after what time $\tau$ the next step of
fork $\nu\in\{L,R\}$ occurs. The formulation via the waiting time density
$\sum_{\mu,\nu}P$ is economical computationally, avoiding a large number
of unsuccessful opening attempts in traditional Langevin simulations. Using
(\ref{gill}) we obtain the single bubble time series in Fig.~\ref{signal}.

\emph{Phage T7 analysis.}~By ME and stochastic simulation we investigate
the promoter sequence of the T7 phage,
\begin{equation}
\begin{array}{l}
\mbox{\small\texttt{\textcolor{white}{AAAA}1\textcolor{white}{%
AAAAAAAAAAAAAAAAAA}20}}\\[-0.15cm]
\mbox{\small\texttt{\textcolor{white}{AAAA}|\textcolor{white}{%
AAAAAAAAAAAAAAAAAA}|\textcolor{white}{AAAAAAAAA}\textcolor{white}{AAA}}}\\[-0.15cm]
\mbox{\small\texttt{5'-aTGACCAGTTGAAGGACTGGAAGTAATACGACTC}}\\
\mbox{\small\texttt{\textcolor{white}{AAA}AG}\textcolor{red}{
\texttt{TATA}}\texttt{GGGACAATGCTTAAGGTCGCTCTCTAGGAg-3'}}\\[-0.15cm]
\mbox{\small\texttt{\textcolor{white}{AAAAAAA}|\textcolor{white}{AA}|
\textcolor{white}{AAAAAAAAAAAAAAAAAAAAAAAAA}|\textcolor{white}{AAA}}}\\[-0.15cm]
\mbox{\small\texttt{\textcolor{white}{AAAAAAA}\textcolor{red}{38}%
\textcolor{white}{A}\textcolor{blue}{41}\textcolor{white}{%
AAAAAAAAAAAAAAAAAAAAAAAAA}68\textcolor{white}{AAA}}}
\end{array}
\end{equation}
whose TATA motif is marked red \cite{kalo}. Fig.~\ref{signal} shows the time
series of $I(t)$ at $37^{\circ}$C for the tag positions $x_T=38$ in the core
of TATA, and $x_T=41$ at the second GC bp after TATA: Bubble events occur
much more frequently in TATA (AT/TA bps are particularly weak \cite{FK}).
This is quantified by the density of waiting times $\psi(\tau)$ in the $I=0$
state, whose characteristic time scale $\tau'$ is more than an order of
magnitude longer at $x_T=41$. In contrast, we observe similar behaviour for the
density $\phi(\tau)$ in the $I=1$ state for $x_T=38$ and 41. Both $\psi(\tau)$
and $\phi(\tau)$ decay exponentially for long $t$; the overlaid lines represent
numerical evaluation of the ME, see \cite{unp}. As shown in
the bottom for the parameters from \cite{FK}, the variation of the
mean correlation time $\tau_{\mathrm{corr}}=\int A(x_T,t)dt$ obtained from
the ME is small for the entire sequence, consistent with the low sequence
sensitivity of $\phi(\tau)$. Note the even smaller variation predicted
for the parameters of \cite{blake}.

\begin{figure}
\begin{center}
\includegraphics[width=8.8cm]{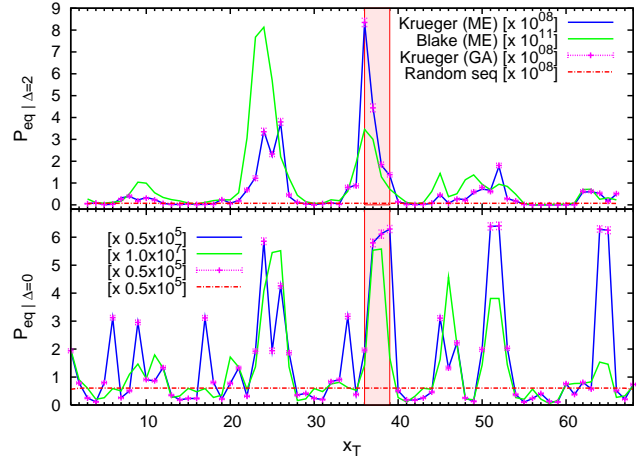}
\end{center}
\caption{Probability to have bps $[x_T-\Delta,x_T+\Delta]$ open.}
\label{open_symm}
\end{figure}

Fig.~\ref{open_symm} shows the equilibrium probability that the bps $[x_T
-\Delta,x_T+\Delta]$ are open, as necessary for fluorescence to occur. We
plot data obtained from the zeroth mode of the ME together with the time
average from the Gillespie algorithm (GA), finding excellent agreement.
Whereas for $\Delta=0$ several segments show increased tendency to
opening, for the case $\Delta=2$, one major peak is observed; the
data from \cite{FK} coincide precisely with TATA, while the data from
\cite{blake} peak upstream. Also shown is a
comparison to the opening probability of a random sequence demonstrating that
the enhanced opening probability at TATA is significant, compare \cite{unp}.
Analysis for various $\Delta$ indicate best
discrimination of the TATA sequence being open for $\Delta=2$. For future
FCS or energy transfer experiments, it therefore appears important to
optimise the $\Delta$-dependence for best resolution, e.g., by adjusting
the linker lengths of fluorophore and quencher \cite{NB}.

\begin{figure}
\begin{center}
\includegraphics[width=8.8cm]{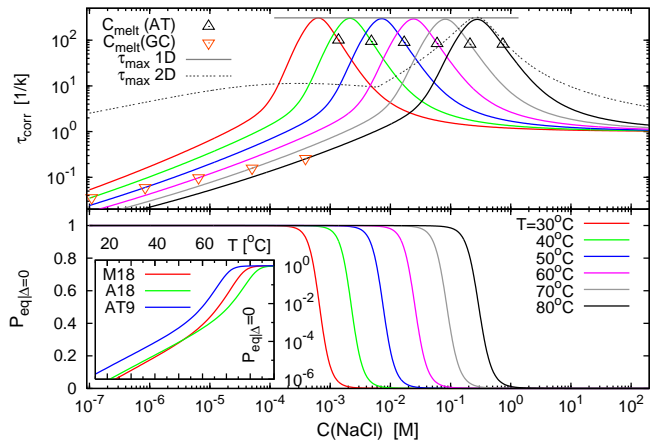}
\end{center}
\caption{Mean correlation time versus salt for various $T$ (top), and melting
curves versus salt concentration and $T$ (bottom).}
\label{salttemp}
\end{figure}

\emph{Nanosensing.}~Fig.~\ref{salttemp} shows the dependence of the mean
correlation time of the AT9 sequence on salt concentration $C$ and $T$. The
variation with $C$ and $T$ is significant, pointing toward potential
applications of DNA fluorescence constructs as nanosensors \cite{nanosens}.
The triangles denote the melting concentration of infinitely long random AT
and GC stretches, respectively (see \cite{FK}). The maxima of the
$\tau_{\mathrm{corr}}$ curves hallmark the critical slowing down of the
autocorrelation at the phase transition point beyond which the bubble is
preferentially open, see also \cite{bicout}. Note that the maxima
coincide with the melting concentrations in the bottom panel. The dashed
line ($\tau_{\mathrm{max}}$ 2D) corresponds to the longest relaxation time
obtained numerically from the ME; it agrees well with $\tau_{\mathrm{corr}}$
close to the maximum, analogously for the other $T$. The horizontal line
($\tau_{\mathrm{max}}$ 1D) represents the longest relaxation time $(2M+1)^2/
\pi^2 k^{-1}$ obtained from the homopolymer model of Ref.~\cite{tobias}
in the limit $u\to 1$, $\sigma_0\to0$ and $c=1$ ($M=27$, length of the
AT9 construct), with the same scaling as the first exit of unbiased
diffusion.

\emph{Discussion.}~Previous bulk melting studies provided DNA stability
data \cite{FK,blake}, on whose basis the relation between local sequence
stability and coding properties of the associated genes was shown
\cite{yeramian,carlon}. However, it is single
molecule experiments that permit to study the \emph{dynamics\/}
of DNA denaturation and renaturation \cite{oleg}. We here derive a physical
framework for the opening and closing fluctuations of intermittent
DNA bubbles in an arbitrary sequence of bps
using the position of the two bubble zipper forks as fundamental coordinates.
By comparison with previously unpublished FCS data we
prove the predictive power of our model.
As complementary approach based on the same (un)zipping rates, we introduced
the stochastic Gillespie simulation, that provides the time series of single
bubble fluctuations. The time averages
from the stochastic simulation agree well with the ensemble properties derived
from the ME. By its computationally attractive
formulation based on the waiting time the Gillespie approach
allows to include additional effects such as protein binding dynamics,
or to consider longer chains and multibubble states. For a long
homopolymer our model is analytically tractable \cite{tobias,unp}.

We used recent DNA stability data from \cite{FK} based on
separation of hydrogen bond and stacking energies, a distinct feature
being the low stacking in a TA/AT stack,
translating into a pronounced instability of the TATA motif, as shown
for the T7 promoter sequence. The relevance of stacking interactions
is also shown in the inset in Fig.~\ref{salttemp} exhibiting
pronouncedly different melting behaviour despite identical AT and GC
contents for the constructs in \cite{oleg,rusu}. Regarding the
biological relevance of TATA, from our analysis it may be speculated
that it is not primarily the bubble lifetime (typically shorter than the
timescale of protein conformational changes) but the
recurrence frequency of bubble events that triggers the initiation of
transcription.
Note that typical binding energies of TATA binding proteins exceed
the free energy to break up TATA, while both energies are comparable
for a random sequence of the same length.

Given the high
sensitivity of bubble dynamics to the stability parameters it should be of
interest to employ FCS on designed DNA constructs to more accurately
obtain stability data for different DNA structures and to calibrate the
(un)zipping rates. 

We thank G.~Altan-Bonnet and A.~Libchaber for sharing the data for
Fig.~\ref{autocorr}, M.~Frank-Kamenetskii for discussion and access to
the new stability data prior to publication, and M. A. Lomholt and
K.~Splitorff for discussion.


\begin{thebibliography}{99}
\bibitem{REM} Due to intrachain coupling (e.g., Rouse), larger bubbles may
involve an additional `hook factor' $m^{-\mu}$ \cite{tobias}.
\bibitem{remm} Except for Fig.~\ref{salttemp}, the parameters are for 100mM
NaCl.
\bibitem{NB} Note that,
assuming a circular bubble of 5 open bps with bp-bp distance 3.4{\AA},
the distance between fluorophore and quencher on bubble opening increases
by 6.5{\AA}, the same magnitude as the F{\"o}rster transfer radius.
\end{thebibliography}
\end{document}